\documentstyle[prb,aps,tighten,multicol]{revtex} 

\def\Journal#1#2#3#4{{#1} {\bf #2}, #3 (#4)}

\begin{document}

\draft 
 
\title{Universal Properties in Low Dimensional Fermionic Systems and Bosonization}

\author{L. E. Oxman$^a$,  D. G. Barci$^{a,b}$ and E. R. Mucciolo$^c$}

\address{a) Instituto de F\'{\i}sica, DFT, 
Universidade do Estado do Rio de Janeiro\\
Rua S{\~a}o Francisco Xavier, 524, 20550-013, Rio de Janeiro, Brazil} 

\address{b) Department of Physics, University of Illinois at Urbana-Champaign\\
1110, W. Green St., Urbana, IL 61801-3080, USA}

\address{c) Departamento de F\'{\i}sica, Pontif\'{\i}cia Universidade
Cat\'olica do Rio de Janeiro,\\ Caixa Postal 38071, 22452-970 Rio de
Janeiro, RJ, Brazil}

\date{December 15, 2000}

\maketitle


\begin{abstract}
We analyze the universal transport behavior in 1D and 2D fermionic systems by following
the unified framework provided by bosonization. The role played by the adiabatic transition
between {\it interacting} and {\it noninteracting} regions is emphasized.
\end{abstract}

\begin{multicols}{2} 
 
\narrowtext 

It is well known that when a system containing complicated interactions shows universal transport behavior, 
this is certainly due to some underlying symmetry principle or topological structure. On the other hand, in 
one-dimensional systems such as quantum wires, in order to understand the universal Landauer conductance, it is important to take into account the adiabatic transition between the interacting Luttinger liquid in the wire and the noninteracting Fermi liquid in the leads.\cite{MPS}  Indeed, the observed deviations of Landauer's conductance from its perfect value $e^2/h$ could be associated to a nonadiabaticity in the experimental 
setup.\cite{TY} 
In this note, we will comment briefly on these topics, that is, universal behavior, symmetries, topology and adiabaticity, by following the unified framework provided by the bosonization of low-dimensional systems.

By now, it is known~\cite{FSBQM} that the bosonization of current correlation functions associated with the free fermion partiton function $Z_0[s]$ can be implemented in terms of an action $K_B[\lambda]$, where
$\exp{i K_B[\lambda]}=\int {\cal D}b\, Z_0[b]\, \exp{i \int d^\nu x\, b\varepsilon\partial\lambda}$, and the {\it exact} 
current bosonization rule $j\leftrightarrow \varepsilon \partial \lambda$. For space-time dimension $\nu =2$, $\lambda$ is a scalar field $\phi$ and $j^\mu\leftrightarrow \varepsilon^{\mu \nu} \partial_ \nu \phi$, while for $\nu=3$, $\lambda$ is a vector field $A$ and $j^\mu\leftrightarrow \varepsilon^{\mu \nu \rho} \partial_ \nu A_ \rho$.
These rules are {\it universal} in the sense that, when a current interaction $I[j]$ is introduced, the following bosonizing mapping can be considered~\cite{univ1}, $K_F[\psi]+I[j]-\int d^\nu x\,  s. j \leftrightarrow 
K_B[\lambda]+I[\varepsilon\partial\lambda]-\int d^\nu x\,  s\varepsilon\partial\lambda$.

For 1D massless Dirac fermions, $K_F[\psi]=\int d^2 x\, i\bar\psi \partial \!\!\!/ \psi$, the fermionic partition function is just a Fujikawa anomalous jacobian and the computed bosonized action corresponds to the usual one,
$K_B[\phi]=\int d^2 x\, \frac{\pi}{2}\, \partial_\mu \phi \partial^\mu \phi$.
In two dimensions, this bosonization procedure is suitable for studying parity breaking systems, such as massive Dirac fermions and nonrelativistic electrons in the presence of a magnetic field. 
We will call ``Perfect Hall Regions'' those regions where the parity breaking parameter is large. There, 
the bosonized action for 2D relativistic large mass Dirac fermions ($m\to \infty$)
turns out to be $K_B[A]\sim K_{\infty}[A]=4\pi S_{CS} [A]$~\cite{FSBQM},
while the bosonized action for 2D nonrelativistic electrons in the presence of a large magnetic field (large $B_{\rm ext}/m$), with filling factor $1$, is
$K_B[A]\sim K_{\infty}[A] = 2\pi S_{CS} [A]+
2\pi \int d^3x\, \{A_0 B_{\rm ext} - \frac{3}{2m}(\vec\nabla\times\vec A)^2 \}.$\vspace{.03in}

\noindent
$\bullet$~\underline{{\bf 1D Systems:}} To study transport through a localized interacting $1D$ 
region $\Omega$ connected to noninteracting Fermi liquid leads, we can model the one channel leads by means of massless Dirac fermions. If the interaction is chiral symmetric and the transition between interacting and noninteracting regions is adiabatic, then an anomalous
chiral current divergence can be derived from the effective bosonized equations of motion,
\begin{equation}
\partial_\mu \left[ \partial^\mu \phi+j^\mu_{loc}
\right] = - (1/\pi)\, E({\bf x},t), 
\label{adiv}
\end{equation}
where $j^\mu_ {loc}$ is a term localized in $\Omega$, and related to the interactions.
This is indeed the case when: i) a current interaction $I[j^\mu]$ is present; ii) a {\it finite} incommensurate Peierls-Fr\"ohlich system (ICDW), at low temperatures, is considered. In i), because of the above mentioned universal current bosonization rule, $j^\mu_{loc}$ turns out to be $1/\pi\, \varepsilon^{\mu \nu} \delta I/\delta j^\nu$, while in the case ii), $j^\mu_ {loc}$ receives the contribution of the localized lattice degrees of feedom.\cite{univ2} We recall that the ICDW system is modelled in terms of a field theory displaying an (anomalous) chiral symmetry~\cite{I} $\psi_R \rightarrow e^{i\pi\alpha} \psi_R~~,~~\psi_L \rightarrow e^{-i\pi \alpha} \psi_L~~,~~\phi\rightarrow \phi -\alpha~~, ~~\Delta \rightarrow e^{i2\pi\alpha}\Delta$, where $\Delta$ is a 
field associated to the lattice displacements ${\cal R}  \left[\Delta \exp{i2k_F x}\right]$. In both cases, from the moment an electric field is switched on until it saturates in a stationary value ${\bf E}(x)$, the asymptotic behavior of the fields corresponds to a motion along the ``direction'' of the chiral  symmetry, as this minimizes the generated action. For instance, in ii) the asymptotic behavior is $\phi=f({\bf x})-kt$, $\Delta = e^{i2\pi kt} D(\bf x)$.\cite{univ2} This means that, in the asymptotic regime, a current $I\leftrightarrow -\partial_0 \phi= k$ and a localized CDW, ${\cal R}  \left[D \exp{i2(k_F x+\pi kt)}\right]$, are generated. Eq.~(\ref{adiv}) becomes
stationary in both cases, and after integrating between two points ${\bf a}$ and ${\bf b}$ on the left (resp. right)
lead, the part depending on the localized term $j_{loc}$ vanishes. Following a Maslov and Stone's reasoning~\cite{MPS}, we have shown~\cite{univ2} that the asymptotic current is indeed $I=k=\frac{1}{2\pi}[V({\bf b})-V({\bf a})]$. In ii), this corresponds to a perfect conductance $e^2/h$ coming from the perfect matching between the Fermi liquid in the leads and the generated CDW in the lattice. 

We remark that the ideal transport in ICDW systems can be generalized to analogous 2D systems.\cite{W} In the following we comment instead on an extension of the previous methods to 2D {\it parity breaking} systems.\vspace{.03in}

\noindent
$\bullet$~\underline{{\bf 2D Systems:}} We suppose here that the
``Hall System'' is perfect everywhere but on a region $\Omega$, where smooth and small
deviations from the Perfect Hall condition are present, that is, in $\Omega$: i) the fermion mass is not so large 
(relativistic fermions), ii) there are deviations from the large magnetic field $B_{ext}$ (that was adjusted to filling factor $1$). Therefore, the free fermion bosonized action will no longer be $K_{\infty}[A]$ (see the expressions quoted above), but on
general grounds it will be gauge invariant and will have the form
$K_B[A]= K_{\infty} + R[\varepsilon\partial A]$, where $R[\varepsilon\partial A]$ is some functional localized in $\Omega$.\cite{univ3} 

In the simpler case i), if current interactions are introduced,
the universal character of the current bosonization rule implies $S_{bos}[A]=4\pi S_{CS} [A] + R[\varepsilon\partial A] + I[\varepsilon \partial A]-\int d^3x s\varepsilon \partial A$, leading
to the equation,
\begin{equation}
4\pi \varepsilon^{\mu \nu \rho} \partial_{\nu} A_\rho - \varepsilon^{\mu \nu \rho} 
\partial_{\nu} [\delta R/\delta j^{\rho}+\delta I/\delta j^{\rho}]
=-\varepsilon^{\mu \nu \rho} \partial_{\nu} s_\rho.
\label{mov}
\end{equation}
When $I[j]$ is also localized in $\Omega$, we have shown~\cite{univ3} that, considering stationary external sources, taking a spatial component in Eq.~(\ref{mov}), and integrating on a line between two points 
${\bf a}$ and ${\bf b}$ placed on different Perfect Hall regions, the part that depends on the localized functionals $R$ and $I$ does not contribute, and we are left with $\left (\int dx^i  \partial_i A_0 \right) =
\frac{1}{4\pi} [V({\bf b})-V({\bf a})]$~\cite{univ3}. The left hand side here is the bosonized expression for the current across the line.
This result is universal and corresponds to a {\it half} perfect transverse conductance.
A similar reasoning can be done in ii), where the transverse conductance $e^2/h$ is seen to be universal,
not depending on the charge density interactions localized in $\Omega$, nor on the particular geometry of this region.\cite{univ3}

Regarding these results, in the case ii), two remarks are in order. Firstly, when the system is 
tunned to filling factor $1$, Pauli's principle turns the interactions irrelevant. However, when small deviations in $B_{ext}$ are present, it is natural to consider interactions localized in $\Omega$ as well. Secondly, we 
recall that two pictures have been risen regarding the question of how the current injected into the sample is distributed there: the edge and bulk current pictures, depending on wether the Hall voltage drops near
the edges or gradually across the sample (see~\cite{T} and references therein). Associated arguments for universality are well known and have been discussed by others. An argument including 
electron-electron interactions has been derived for the first picture~\cite{F}, by means of an (edge) current 
algebra analysis in an effective Thirring model defined on the sample's boundary. 
However, in Hall experiments using Corbino geometry, where the boundaries are not relevant, all the current must run through the bulk.\cite{V} In this case, the universal Hall conductance has been derived, for an interacting many-body (gapped) system defined on a torus, by expressing it as a topological Chern invariant.\cite{niu}
We see that the alternative derivation we have analyzed here also corresponds to a {\it bulk} universality, including electron-electron interaction effects where small magnetic field deviations are present.\vspace{.03in}

In summary, symmetry and topological properties are stressed by the bosonization mapping, while universal information is already present among its rules ($j\leftrightarrow \varepsilon \partial \lambda$). 
In order for these properties lead to universal transport behavior, the adiabatic transition between the interacting and noniteracting regions plays a fundamental role, enabling a unified treatment of both regions in a single theory, and establishing a particular matching between them. Whenever these conditions are met, strong constraints are imposed on transport in 1D or 2D systems: 
the longitudinal and transverse conductance become dominated by the Fermi liquid in the leads
or the Perfect Hall regions.

\acknowledgments

We are indebted to S.P. Sorella, I.V. Krive and C.D. Fosco for useful discussions.
The Funda{\c {c}}{\~{a}}o de Amparo {\`{a}} Pesquisa do Estado do Rio de Janeiro (FAPERJ), 
the CNPq-Brazil, and the SR2-UERJ are acknowledged for the financial support.
D.G.B. is partially supported by the NSF grant
No. DMR98-17941, UERJ and by the CNPq through a postdoctoral fellowship.

\end{multicols}

\end{document}